\documentclass[aip,graphicx,reprint]{revtex4-1}
\usepackage[usenames,dvipsnames]{xcolor}   
\usepackage{graphicx}
\usepackage{amsmath}
\usepackage{url}
\DeclareMathOperator*{\argmin}{arg\,min}

\begin{document}
\title{Embedding quantum statistical excitations in a classical force field}
\author{Susan R. Atlas}
\email[]{susier@unm.edu}
\affiliation{Department of Chemistry and Chemical Biology, Department of Physics and Astronomy, and Center for Quantum Information and Control, University of New Mexico, Albuquerque, NM 87131}
\date{\today}

\begin{abstract}
Quantum-mechanically-driven charge polarization and charge transfer are ubiquitous in biomolecular systems, controlling reaction rates, allosteric interactions, ligand-protein binding, membrane transport, and dynamically-driven structural transformations. Molecular dynamics (MD) simulations of these processes require quantum mechanical (QM) information in order to accurately describe their reactive dynamics.  However, current  techniques---empirical force fields, subsystem approaches, {\it ab initio} MD, and machine learning---vary in their ability to achieve a consistent chemical description across multiple atom types, and at scale.  Here we present a physics-based, atomistic force field, the \textit{ensemble DFT charge-transfer embedded-atom method}, in which QM forces are described at a uniform level of theory across all atoms, avoiding the need for explicit solution of the Schr\"{o}dinger equation or large, precomputed training datasets. Coupling between the electronic and atomistic length scales is effected through an ensemble density functional theory formulation of the embedded atom method originally developed for elemental materials. Charge transfer is expressed in terms of ensembles of \textit{ionic} states basis densities of individual atoms, and charge polarization, in terms of atomic \textit{excited} state basis densities. This provides a highly compact yet general representation of the force field, encompassing both local and system-wide effects. Charge rearrangement is realized through the evolution of ensemble weights, adjusted at each dynamical timestep via chemical potential equalization.
\end{abstract}

\maketitle
\section{Introduction}
Classical force fields are essential tools for the atomistic simulation of biomolecular systems. They can be used to probe  the structure and biochemical interactions of peptides, small organic molecules, proteins and their ligands, nucleic acids, and larger structures such as lipid membranes and ion channels. The most widely-used empirical force fields are those based on the CHARMM \cite{mackerell1998,brooks2009,vanommeslaeghe2010}, AMBER\cite{cornell1995,wang2004,maier2015} and OPLS-AA\cite{jorgensen1996} families. Solvation effects---critical to an accurate description of protein structure, dynamics, and function\cite{bellissent2016}---are described via implicit or explicit (atomic) water models,\cite{guillot2002,vega2011,onufriev2018} the latter often employing fixed charges, as in the widely-used TIP \cite{jorgensen1983} and SPC\cite{berendsen1987} models.  The additive energy function describing the interactions among atoms  includes harmonic bond and angle terms, bonded torsional energies, and electrostatic and non-bonded (van der Waals) interactions, with fixed, atom-centered partial charges derived from electrostatic models or quantum mechanical calculations.\cite{jorgensen2005} This general class of empirical force fields is also used as a jumping-off point for developing polarizable force fields\cite{demerdash2018,jing2019} and other variants,\cite{nerenberg2018} and are regularly revisited, refined, reparameterized, and extended to encompass new classes of biomolecules and their interactions.\cite{lindorff2010,vanommeslaeghe2012a,nerenberg2018,he2020}  Efforts are underway to extend force fields and associated water models\cite{shabane2019} to describe increasingly complex biochemical and biophysical processes, such as the fluctuations of intrinsically-disordered proteins among multiple structural minima,\cite{wang2017,huang2017,robustelli2018} and the impact of  ionic interactions on nucleic acid catalysis,\cite{yang2006} RNA folding, \cite{Nguyen2020} and the mechanochemistry and catalytic cycles of molecular motor proteins\cite{kiani2016} such as kinesin \cite{Parke2010,hariharan2009,hwang2008,cross2016} and myosin.\cite{lawson2004,ovchinnikov2010,li2013}  
While considerable biophysical and structural insights can be gained using all-atom techniques based on empirical classical force fields, in many instances a quantum-level description of atomic interactions is required.  The atomistic ReaxFF force field\cite{chenoweth2008,senftle2016} reframes the energy function in terms of dynamically-computed bond order and charge polarization contributions. However, due to its empirical design, and the intrinsically complex nature of reactive chemistry, parameter transferability can be limited even between nominally similar problem subclasses.\cite{senftle2016}  Fragmentation methods,\cite{gordon2012,pruitt2014} which partition a system into interacting atom-groupings, each of which is treated quantum-mechanically, are a promising approach for modeling biochemical systems at scale, but require coarse-grained partitioning into functional groups. The pioneering  QM/MM\cite{warshel1976,field1990} method can be used to focus attention on a smaller, predefined, region using quantum mechanics, while treating the rest of the system classically. This technique requires careful consideration of boundary and embedding effects, including interactions between the QM and MM regions, but is a valuable tool for studying free energies, catalysis\cite{van2013} including ATP hydrolysis,\cite{mcgrath2013,kiani2014,kiani2015,baldo2020} and photochemistry.\cite{boulanger2018}. {\it Ab initio} molecular dynamics\cite{car1985,carloni2002} integrates the quantum mechanical evaluation of the forces on each atom---typically done using DFT---with the classical dynamical evolution of the system itself. However, since quantum calculations are performed as part of the simulation, this method is limited in the size of the systems (hundreds of atoms) and timescale of simulations (${\cal O}(100$~ps)) that can be studied.\cite{zhang2018}

An alternative, and increasingly popular, approach is to utilize machine-learned representations of the force field. This is done by constructing neural network classifiers\cite{behler2007,behler2011} for predicting individual atomic contributions to the overall energy function: the parameters of the neural network are determined by training the predictions of the force field energy against a pre-computed, quantum mechanical database.  At a given timestep in a simulation, the input to each atom's classifier is a set of symmetry-respecting \textit{features} (also known as {\it descriptors}) characterizing the local conformational and chemical environment. Examples include the ANI-1 potential for small organic molecules,\cite{smith2017} and the DeepMD potential, in which symmetry functions with a pre-defined form are replaced by a local coordinate representation.\cite{han2018,zhang2018} 

In all of these methods, there is an inevitable tradeoff between system size, diversity of atomic composition, and level of quantum mechanical description that can be modeled. This limits current ability to describe nonlocal, quantum-mediated effects in large systems---for instance, allosteric protein interactions and charge-transfer-driven structural changes,\cite{Wei2016} or biochemical processes such as motor protein procession along a microtubule in a cell,\cite{hancock2016} which are driven by the coupling of chemical and mechanical transitions occurring over long (millisecond to second) time scales.  For such problems, chemically-accurate atomistic potentials are also needed in order to identify the relevant substructures---switches, salt bridges, disordered regions---and Markov states\cite{noe2017} for constructing atomistically-informed multiscale kinetic models incorporating an accurate description of the underlying chemistry.\cite{jacobson2014SA,jacobson2014load,jacobson2017}

Here, we present a physics-based approach for holistically incorporating quantum effects within an atomistic force field. This is accomplished by expressing the potential functional in terms of dynamically-evolving atom-in-molecule electron densities,\cite{Atlas2020} which are further represented as statistical ensembles of isolated neutral atom, excited state, and ionic state densities.  These {\it basis densities} serve as fundamental constructs of the model.\cite{Amokwao2020} Since the basis densities are those of isolated atoms, they can be precomputed at an arbitrarily high level of theory, including spin polarization, electron correlation, and even relativistic effects in the case of  heavier atoms. The framework we present here---a formulation of the charge-transfer embedded-atom method (CT-EAM)\cite{valone2006Kos,muralidharan2007ED} to facilitate large-scale simulations of biomolecular systems---has its roots in the empirical embedded-atom method (EAM) for metallic \cite{daw1983,daw1984,foiles1986} and covalent\cite{baskes1987,baskes1989} systems, and the density functional theory (DFT) of quantum systems as elucidated by Hohenberg and Kohn\cite{hohenberg1964} and Kohn and Sham,\cite{kohn1965} with subsequent extensions to ensembles of atomic charge states\cite{perdew1982} and ensembles of atomic excited states.\cite{kohn1986,gross1988var,gross1988DFT,oliveira1988} We derive a general force field expression---the ensemble DFT charge-transfer EAM (CT-EAM)---for coupling quantum mechanical densities to atomistic energies, including both local and global correlations.  The quantum mechanical coupling is effected through the medium of the electron density, a physical observable, in order to flexibly and consistently describe reactive chemistry, without local bias, across large biomolecular systems.

\section{Theoretical Background and Methods}
This section provides the theoretical and conceptual background for establishing systematic linkages between the electronic and atomistic length scales through DFT and the DFT-based EAM model. EAM was originally developed for simulating elemental materials systems,\cite{daw1983,daw1984,foiles1986,baskes1987,baskes1989} and is extended here to incorporate charge polarization and charge transfer effects in molecular systems.  We begin by reviewing the related valence bond, fluctuating Hamiltonian (FH)  model\cite{valone2011IM,valone2011WF,valone2014} to highlight the essential role played by an ensemble atom-in-molecule picture in both wavefunction- and density-based formulations of a general charge-transfer EAM model.\cite{valone2006Kos}  Energy-density duality is presented as a unifying concept for understanding the tradeoffs involved in constructing a DFT-based force field.  We then describe the specific energy and density foundations of the ensemble DFT CT-EAM potential: the EAM method itself, which provides the DFT-based scaffold incorporating charge polarization and charge transfer effects within ensemble DFT CT-EAM; and a density functional theory of the atom-in-molecule, couched in terms of statistical ensembles of atomic charge and excited states. 
\subsection{Valence bond origins: spectral representation and the atom-in-molecule} 
Potential models based on a valence bond wavefunction  picture\cite{aaqvist1993,schmitt1998,morales2001,morales2004,valone2004,valone2008} have long provided important insights into the interplay between energy, charge, and chemical potential for accurately describing bond formation and breaking in charge-transfer systems, even at the level of diatomic models.\cite{valone2006Kos,valone2011IM}  

Using a simple two-state valence bond model, Valone and Atlas\cite{valone2006ED} showed that by applying two key concepts---(i) transformation from a valence-bond picture (with relative ionic and covalent contributions characterized by an ionicity $\gamma$) to a spectral representation; and (ii) the introduction of an atom-in-molecule decomposition characterized by a charge transfer $q$---it was possible to extend the the ensemble model of charge transfer of Perdew {\it et al.}\cite{perdew1982} from weakly-interacting to strongly-interacting subsystems.

Starting from these ideas, Valone proposed and developed the fluctuating Hamiltonian (FH) model\cite{valone2011IM,valone2011WF,valone2014,valone2016} which uses an ensemble wavefunction picture and an explicit Hamiltonian formulation to describe fluctuating charge states of an atom interacting with its environment.  The model yields generalizations of familiar chemical concepts---Mulliken electronegativity, Parr-Pearson hardness, ionization potential, electron affinity, and charge-transfer gap energy---expressed as combinations of Hamiltonian matrix elements and density matrix elements, together with an expression for the energy of an atomic fragment, incorporating charge-transfer dependence through the ionicity. Furthermore, by mapping the charge-state-dependent fluctuating Hamiltonian picture onto Moffitt's Hamiltonian representation for the atom-in-molecule,\cite{moffitt1951} a charge-transfer EAM embedding energy\cite{valone2006Kos} can be defined.\cite{valone2014,valone2016}  The remaining two-body terms in the atom-in-molecule energy expression correspond to electrostatic interactions, as in the original EAM ({\it cf.}~Eq.~(\ref{eq:EAM}) below).\cite{daw1983,daw1984}  The FH model thus integrates familiar chemical concepts within an atom-in-molecule EAM framework, using wavefunction-based reasoning.   

\subsection{Energy-density duality}
That an intimate relationship exists between the total energy of a many-electron system and its spatial electron density distribution $\rho({\bf r})$ was postulated almost a century ago, as what is now known as Thomas-Fermi (TF) theory.\cite{lieb1977}  The TF picture of the electron cloud in an atom is that of an inhomogeneous electron gas, whose spatial distribution can be described by a mean field theory. The kinetic energy is approximated using a local density approximation\cite{hohenberg1964}, i.e., the local application of the expression for the kinetic energy of a homogeneous electron gas, and the total energy is given by:\cite{lundqvist1983}
\begin{equation}
    E_{\rm TF}[\rho({\bf r})] = T_{\rm TF}[\rho({\bf r})] + U_{\rm H}[\rho({\bf r})]  + \int \rho ({\bf r})\, v({\bf r})\, d{\bf r},
    \label{eq:TF}
\end{equation}
where $T_{\rm TF} [\rho({\bf r})] = C_{\rm K} \int [\rho({\bf r})]^{5/3} d{\bf r}$ ($C_K$ is a constant),
$U_{\rm H}[\rho({\bf r})] = 1/2 \int \int d{\bf r}\, d{\bf r}' \rho({\bf r})\, \rho({\bf r}')/|{\bf r} - {\bf r}'|$ is the classical Hartree electrostatic energy, and $v({\bf r})$ is the external potential due to the positive nuclear charge of the atom.

Some forty years later, this appealing picture found formal expression in the celebrated theorem of Hohenberg and Kohn (HK),\cite{hohenberg1964} which established the foundation for modern density functional theory (DFT).  The proof of the Kohn-Sham (KS) theorem and coupled single-particle equations followed one year later.\cite{kohn1965} The KS equations provide a Hartree-Fock-like framework for the practical application of the HK theorem to electronic structure calculations, including statistical and spatial electron correlation,\cite{kohn1965} and the relative simplicity of their application has transformed quantum chemical calculation of molecular systems.  In this discussion, we focus exclusively on the conceptual implications of the HK theorem, since our objective is to derive a purely density-dependent force field, without reliance on wavefunction theory or Kohn-Sham orbitals.\cite{kohn1965}

The Hohenberg-Kohn theorem consists of three parts:\cite{dreizler1990} (i) the statement that knowledge of the ground state electron density $\rho({\bf r})$ for a many-electron system uniquely determines the configuration (magnitude and locations of nuclear charges---equivalently, the external potential $v({\bf r})$) for the interacting atoms that gave rise to $\rho({\bf r})$; (ii) formal proof of the existence of an energy functional $E_v[\rho({\bf r})]$ that provides an exact and complete representation of the quantum mechanics giving rise to the system ground state; and (iii) a corresponding density-based variational principle for determining the ground state density $\rho_{gs}({\bf r})$ by minimizing $E_v[\rho({\bf r})]$ over the space of all possible densities.

Taken together, statements (ii) and (iii) imply that the solution of the many-electron Schr\"{o}dinger equation, including a complete description of quantum mechanical exchange and correlation, can be replaced by the variational minimization of a simple energy functional, depending only on the electron density and external potential $v({\bf r})$ of the system:  
\begin{equation}
  E_v[\rho({\bf r})] = F_{\rm HK}[\rho({\bf r})] + \int \rho({\bf r})\, v({\bf r})\, d{\bf r},
  \label{eq:HK}
\end{equation}
with 
\begin{equation}
   \rho_{gs}({\bf r}) = \argmin_{\rho({\bf r})} E_v[\rho({\bf r})].
\end{equation}
$v({\bf r})$, the system-specific external potential due to all $N_A$ atoms in the system, is given by:
\begin{equation}
  v({\bf r}) = -\sum_{i=1}^{N_A} \frac{Z_i}{|{\bf r} - {\bf R}_i|},
\end{equation}
and $F_{\rm HK}[\rho({\bf r})]$ is the {\it universal} energy density functional. $F_{\rm HK}[\rho({\bf r})]$ consists of the sum of the total kinetic energy and quantum-mechanical electron-electron interaction energies of the system.  Comparing Eqs.~(\ref{eq:TF}) and (\ref{eq:HK}), it is easy to see how the HK energy expression generalizes the original intuition of TF theory.

The functional $F_{\rm HK}[\rho({\bf r})]$ is universal because, unlike $v({\bf r})$, it is system-independent: the same $F_{\rm HK}[\rho({\bf r})]$ applies equally to atoms, molecules, and solids. The other remarkable feature of $F_{\rm HK}[\rho({\bf r}$)] (or equivalently, the exchange-correlation energy functional $E_{\rm xc}[\rho({\bf r})]$ appearing in the KS equations\cite{fn1}), is that, by definition, it embodies the full quantum mechanical complexity of the system, including statistical (exchange) and spatial electron correlations.  These correlations would ordinarily require wavefunction quantum mechanics (e.g., configuration interaction or coupled-cluster solutions of the Schr\"{o}dinger equation to determine the full many-electron wavefunction $\Psi({\bf r}_1, {\bf r}_2, \cdots {\bf r}_N)$), at many orders of magnitude greater computational cost. Instead, these correlations are completely folded into the universal functional expression.  However, since the HK theorem only proves the existence of the universal functional, the functional must be approximated in practice.\cite{cohen2012,sun2015,medvedev2017,williams2020}

It is important to note that the HK theorem not only opened the door to practical, accurate and efficient electronic structure calculations using the Kohn-Sham equations;\cite{kohn1965} it has also led to a fundamental shift in thinking about how to construct models of atomic interactions to directly incorporate quantum mechanical effects without the need to explicitly solve for a many-body wavefunction.  We refer to this profound---and in principle, exact---relationship between the energy and density as \textit{energy-density duality}:\cite{fn2} the energy functional implies the ground state density through the HK theorem variational principle; conversely, the ground state density implies the energy, through the existence of the universal functional.

Energy-density duality has inspired the development of numerous models in which atomic-like densities are used to construct an energy model for a larger system.  Early examples include the Gordon-Kim model of rare gas interactions\cite{gordon1972} and its extensions\cite{lacks1993}; the Harris functional;\cite{jharris1985} Cortona's ``atoms in solids'' method;\cite{cortona1991} frozen DFT;\cite{wesolowski1993,wesolowski2015} and the self-consistent atomic deformation method.\cite{boyer2008} Over the past decade, there has been renewed interest in DFT fragment-based approaches descended from these earlier ideas, including partition DFT,\cite{elliott2010} DFT embedding theory, \cite{huang2011,huang2011a,manby2012} and subsystem DFT\cite{jacob2014,sun2016}. These approaches implement a variety of theoretical techniques to impose a degree of quantum mechanical self-consistency between a smaller, embedded system---an atom, or cluster of atoms---treated at a different level of theory, and its environment.  This enables a subsystem to be studied at a higher level of theory than its surroundings. The methods all require some form of subsystem registration with the computed wavefunction or density of the larger system---often accomplished through a Kohn-Sham potential---and as such, they are primarily aimed at computing the static electronic structures of large systems at fixed geometries.  As with any fragment-based method, the choice of how to define the subsystem and environment can have a significant impact on results.\cite{sun2016}

Since the energy is a functional of the density, the flip side of energy-density duality is to focus attention first on the electron density, in addition to, or lieu of, a combined energy-density model. A notable early effort in this direction---like the present work, statistical in nature, and aimed at computing dynamically-evolving forces for large systems---is the elegant path integral formulation of Pratt and co-workers.\cite{harris1985,pratt1988,hoffman1988,pratt1990}  Harris and Pratt used a discretized propagator method to show\cite{harris1985} that the electron density $\rho({\bf r})$ of a system could be expressed in terms of an effective single-particle potential\cite{fn3} via a $p$-dimensional integral over the coordinates of all possible $p$-segment paths initiating and terminating at ${\bf r} = {\bf r}_0 = {\bf r}_p$. The expression is remarkably simple:\cite{hoffman1988}
\begin{equation}
    \rho({\bf r}) = 2\int d{\bf r}_1 \ldots \int d{\bf r}_{p-1} \left ( \frac{pk_p}{2\pi l_p}\right)^{3p/2} J_{3p/2} (k_pl_p)\, \eta(k_p)^2,
    \label{eq:HP}
\end{equation}
where 
\[
l_p^2 \equiv p \sum_{j=0}^{p-1} ({\bf r}_{j+1} - {\bf r}_j )^2,
\]
and 
\[
\frac{\hbar^2 k_F^2}{2m} = \epsilon_F - \frac{1}{p} \sum_{j=1}^p v({\bf r}_j; p).
\]
Here $\eta$ is the Heaviside function; $J_\nu(z)$ is the Bessel function of order $\nu$, and $\epsilon_F$ is the Fermi energy.\cite{harris1985,hoffman1988} For the case $p = 1$, and suitable choice of $v({\bf r})$, this result reduces to Thomas-Fermi theory; for $p\rightarrow\infty$, it becomes exact.\cite{harris1985}  The expression can thus be seen as interpolating beween the  semi-classical Thomas-Fermi and fully-quantum mechanical regimes. While technically challenging to implement in practice due to the requisite high-dimensional numerical integration,\cite{pratt1988,hoffman1988} the Harris-Pratt formulation represents an important conceptual advance in focusing attention squarely on the electron density as a preferred target for modeling large systems.  In particular, it embodies several features shared by the present work: it is orbital-free; it takes a bottom-up approach to evaluating the electron density scalar field for the complete system; global information from the larger system is incorporated into the evaluation of $\rho({\bf r})$, through contributions to argments in the integrand from the potential along sampled paths in the neighborhood of ${\bf r}$; the expression is statistical in nature; and it trains focus on the electron density as a more compact embodiment of the quantum substructure of a larger system than wavefunction-based approaches.\cite{pratt1990}  
    
As detailed below, the statistical modeling of $\rho({\bf r})$ in the present work is accomplished through an ensemble, rather than path integral, formulation.  Sampling of the local chemical environment is done through an ensemble spectral representation of atom-in-molecule densities\cite{Atlas2020} and EAM energies, and global information about the system is transmitted via  chemical potential equalization adjustment of the corresponding ensemble weights.

\subsection{The embedded-atom method and its density functional foundations}
The embedded-atom method (EAM) for metals,\cite{daw1983,daw1984} the modified EAM (MEAM) for covalent materials,\cite{baskes1987,baskes1989} and their many extensions\cite{fn7} are among the most widely-used interaction potentials for simulating materials structure, defects, and phase diagrams.\cite{lee2010,foiles2012}  The physical picture underlying the EAM {\it ansatz} was motivated by the development of density functional theory,  through the quasiatom theory of Stott and Zaremba\cite{stott1980} and concurrently-developed effective medium theory of Norskov and Lang.\cite{norskov1980} These works analyzed the energy of an atom embedded in a uniform electron gas with compensating positive background. The empirical EAM model was designed to address the more complicated scenario of many interacting atoms, each viewed as an impurity embedded within a locally-defined host (electron density). 

In a seminal work published several years after the introduction of the original EAM, Daw\cite{daw1989} demonstrated that the EAM could be derived---modulo an error term shown to be small under certain well-defined conditions---directly from Hohenberg-Kohn density functional theory. Although not presented in multiscale terms, this work demonstrated a direct link between the quantum and atomistic length scales through the electron density.  It also suggested\cite{valone2006Kos} the tantalizing prospect of developing a theoretically-grounded charge-transfer extension of the EAM based on extensions to DFT itself. Demonstrating that this is indeed possible is the central result of this work.

The EAM expression for the cohesive energy $E_{\rm coh}$ (equivalent to the interaction potential $V(\{{\bf R}_i\})$ for a configuration of atomic nuclei located at $\{{\bf R}_i$\}) is remarkably simple.  It is written as the sum of individual atomic embedding contributions, and a term consisting of pairwise electrostatic interactions between nucleii:
\begin{equation}
E_{{\rm coh}} = \sum_{i=1}^{N_{\rm at}} E_i,
\end{equation}
where 
\begin{equation}
   E_i \equiv F_i(\bar \rho_i(\textbf{R}_i)) + \frac{1}{2} \sum_{j\ne i}
\phi_{ij}(R_{ij}).  \label{eq:EAM}
\end{equation} 
$F_i$ is an element-dependent {\it embedding function} of the effective local background electron density $\overline{\rho}_i({\bf R}_i)$ at atomic site $i$, and represents the collective many-body effects of the other atoms in the host system, and $\phi_{ij}$ is the effective electrostatic pair potential between atoms $i$ and $j$. The inclusion of the many-body term $F_i$ in the energy expression differentiates the EAM from earlier two-body potentials used in materials modeling, including ionic systems.

In practice, the constant background value ${\bar \rho_i}$ at the site of each embedded nucleus must be approximated. In the standard approach, static model electron densities are associated with each nucleus, and parameterized along with the other functions in the model.  $\overline{\rho}_i({\bf R}_i)$ is approximated as the sum of the tails of the $n_i$ nearest-neighbor atom electron densities at site $i$:
\begin{equation}
\overline{\rho}_i({\bf R}_i) \simeq \sum_{\stackrel{j=1}{j\ne
i}}^{n_i}{\rho_j^a({\bf R}_{ij})}. \label{rhobar}
\end{equation}
where $\rho_j^a$ corresponds to the isolated atom electron density of neighbor $j$.  Second-nearest-neighbor methods  have also been developed.\cite{lee2010}

The analysis of Daw showed that for the original EAM, with no charge polarization and no charge transfer, the optimal (energy-minimizing) embedding density could be chosen to minimize the error term resulting from the definition of the embedding term as a difference of DFT energy functionals.\cite{daw1989}  This point will be discussed in greater detail below.  Daw also suggested the possibility of computing the embedding density as a weighted average over the contributions of neighboring atoms within a suitable radius.\cite{daw1989}  It has been noted that different physical interpretations of the background density can lead to very different EAM parameterization schemes, and can also limit the extent to which additional energy scales, associated with charge transfer, can be incorporated into the baseline model.\cite{valone2014}  This is yet another manifestation  of energy-density duality. 

\subsection{Density functional theory of the atom-in-molecule}
The concept of the atom-in-molecule\cite{fn6} has a long history, dating back to the ideas of Lewis,\cite{lewis1916} Moffitt,\cite{moffitt1951} Hirshfeld,\cite{hirshfeld1977} Parr and coworkers,\cite{parr1978,nalewajski2000} and Bader.\cite{bader1985,bader1990} In the modern formulation, one asks how to decompose a given electron density distribution $\rho({\bf r})$ for a molecule or a material into ``chemically-reasonable'' atom-like components $\rho_i^*({\bf r})$:
\begin{equation}
  \rho({\bf r}) = \sum_{i=1}^{N_{at}} \rho_i^*({\bf r}),
\label{rhoMap}
\end{equation}
where $N_{at}$ is the number of atoms in the system, and * is used to denote  atom-in-molecule quantities. In principle, there are an infinity of ways to accomplish the decomposition, and the diversity of available schemas reflect differing theoretical and practical {\it desiderata}.  Since the decomposition can be used to compute an effective charge $q_i$ for the $i$th atom-in-molecule
\begin{equation}
    q_i = Z_i - \int \rho_i^*({\bf r})\, d{\bf r},
\end{equation}
any atom-in-molecule method allows comparison to be made with charge partitioning\cite{lowdin1950,mulliken1955} or other atom-in-molecule decomposition techniques.\cite{bader1990,heidar2017} Effective charges derived from atom-in-molecule decompositions can also be used to compute parametrization data for fixed-charge empirical force fields, or as serve target values for training empirical or machine-learned potentials.\cite{verstraelen2016,nerenberg2018,zubatyuk2019}

Historically, many atom-in-molecule decomposition strategies have been designed with the goal of transferability, e.g., the construction of reusable chemical ``Legos.''\cite{walker1993}  However, since the objective in this work is to forge a systematic theoretical link between the quantum and atomistic length scales, we have developed an approach,\cite{Atlas2020} summarized in the following subsections, that does not impose a transferability bias; instead, the \textit{chemical environment itself} dynamically determines the atoms-in-molecule, consistent with structural and electronic changes occuring both locally and globally. The decomposition is a formal consequence of the Hohenberg-Kohn theorem,\cite{hohenberg1964} and there is no requirement of external theoretical\cite{bader1990} or reference state\cite{parr1978,nalewajski2000,bultinck2007} constraints.  Since it follows directly from the HK theorem, the decomposition  also avoids the need for self-consistent solution of a set of Kohn-Sham equations coupling the atoms-in-molecule to the larger system.\cite{cohen2007,atlas2007} The price paid for this apparent simplicity is that the decomposition, while unique, is not exact.  Nevertheless, it has been shown to reproduce Bader charges as a function of internuclear separation for two challenging chemical systems, LiF and CO,\cite{Atlas2020} a surprising result given the very different nature of the two approaches. Whereas Bader's method\cite{bader1985,bader1990}  partitions a molecular density into topologically-distinct ``basins'' of electronic charge, the density deconstruction procedure allows charge clouds of contributing atoms to overlap and intermix over all space.  The agreement in effective charges derived from the two methods can be seen as reflecting the overriding importance of the external potential---as emphasized by the HK theorem---in determining the fate of electronic charge transfer for a given molecular geometry.

The next section summarizes the two ensemble DFT theories required by the density deconstruction procedure.\cite{Atlas2020}  The first is a formulation of ensemble DFT for excited states,\cite{gross1988var,gross1988DFT,oliveira1988} to enable a description of atomic charge density polarization; the second is Perdew {\it et al.}'s ensemble DFT for fractional charge states,\cite{perdew1982} to enable a description of charge transfer.  Both of these formalisms are valid only for isolated atoms; their application to the interacting atom case is discussed at the end of the section.     

\subsubsection{Ensemble DFT formulations of charge polarization and charge transfer}
{\it Charge polarization}. Gross {\it et al.} proved an extended Rayleigh-Ritz variational principle\cite{gross1988var} for ensembles, and used this result to prove an ensemble generalization of the Hohenberg-Kohn theorem.\cite{gross1988DFT}  Consider an atom A with external nuclear potential $v_{\rm A}({\bf r}) = -Z_{\rm A}/|{\bf R}_{\rm A}  - {\bf r}|$, energy spectrum $\{E_i\}$, and a chosen set of $N_{ens}$ non-zero weights $\{\omega_i\},\ \ i=1,\ldots N_{ens}.$\cite{gross1988DFT} The energies are ordered as $E_1 \le E_2 \cdots E_{N_{ens}}$, and the weights are ordered such that $\omega_1 \ge \omega_2 ... \ge \omega_{N_{ens}} > 0.$  The weights are normalized to 1. Note that the ensemble state of atom A correspnds to an excitation only; the total number of electrons remains constant. In this case,  Gross {\it et al.} showed that the ensemble energy is given by:
\begin{equation}
E_v[\{\omega_i\};\rho({\bf r})] = \sum_{i=1}^{N_{ens}} \omega_i E_i,  
\end{equation}
and the DFT ensemble-energy-minimizing electron density is:
\begin{equation}
    \rho({\bf r};\{\omega_i\}) = \sum_{i=1}^{N_{ens}} \omega_i \rho_i({\bf r}),
\end{equation}
where $\rho_i({\bf r})$ is the electron density of the state with energy $E_i$. This ensemble excited state picture has an alternative interpretation in terms of an electronic canonical (thermal) ensemble.\cite{kohn1986}

{\it Charge transfer.} For fractional charge state ensemble DFT, Perdew {\it et al.} considered an isolated open-system atom with externally-imposed charge $q$ ($0 \le q \le 1$). Using the constrained-search formulation of DFT\cite{levy1979}, they showed that the minimizing energy of the system takes an ensemble form, but with only two integer charge states $M=Z$ and $M+1$ of the atom contributing. The expression for the ensemble energy is:\cite{perdew1982}
\begin{equation}
E_v = (1-q)E_M + qE_{M+1}, 
\label{eq:PPLB-1}
\end{equation}
and the DFT ensemble-energy-minimizing electron density $\rho({\bf r})$ is given by:
\begin{equation}
    \rho({\bf r}) = (1-q)\rho_M({\bf r}) + q\rho_{M+1}({\bf r}).
    \label{eq:PPLB-2}
\end{equation}
Dreizler and Gross have noted that the constrained-search formulation for the isolated atom can be extended to include a manifold of states with arbitrary integer numbers of electrons, although in the isolated atom case the minimum energy is expected to be attained for the two-state case.\cite{dreizler1990} As with the excited state theory, there is an alternative thermodynamic interpretation of this result, in terms of an electronic grand canonical ensemble picture.\cite{gyftopoulos1968,perdew1982}

{\it Strong interactions and the atom-in-molecule.} The excited state and charge state ensemble expressions apply only to an isolated atom experiencing a ``{\it deus ex machina}'' fractional excited state occupation $\omega$ or fractional charge transfer $q$\cite{valone2006ED}---{\it i.e.,} the external imposition of a perturbation on the isolated system.  This restriction is implicit in the assumption of ordered weights and states in the work of Gross {\it et al.}, and in the assumption of a weak interaction between an atom and its charge reservoir, in the work of Perdew {\it et al.}  A general atomistic force field, however, must be able to describe an atom experiencing \textit{significant} charge polarization or charge transfer---including the possibility of fluctuating charge states exceeding unit charge---resulting from chemical bonding or reactive interactions. That is, it must be able to describe an arbitrary atom-in-molecule.

The charge transfer or charge polarization induced in the strong coupling case can be seen as equivalent to imposing an external interaction potential $v_{int}({\bf r})$ on the atom, in addition to its isolated-atom nuclear potential,\cite{cohen2006,cohen2007,scheffler2007} leading to an effective ``external potential'' $\tilde{v}({\bf r}) = v_{\rm A}({\bf r}) + v_{int}({\bf r})$ for the atom-in-molecule. Of course, this potential can be determined only as the result of a self-consistent electronic structure problem for the full system, which, by design, is not possible for a classical potential. Our solution, therefore, is to construct a representation of the atom-in-molecule density that utilizes the isolated atom ensemble states as {\it basis densities}. This representation, while approximate, can be shown to be a unique consequence of the Hohenberg-Kohn theorem,\cite{Atlas2020} and is summarized in the next section.  

The basis density representation implements two changes relative to the weakly-interacting ensemble formalisms: the manifolds of contributing states can now include any physically-accessible states of the isolated atoms or ions; and the ensemble weights are no longer required to be ordered according to energy\cite{gross1988var,gross1988DFT} or charge state.\cite{perdew1982}  The impact of the effective potential is detected indirectly, through modified ensemble weights appearing in both the density and energy contributions to the atomistic model.  In fact, the appearance of charge-dependent weights in an ensemble representation of an atom-in-molecule can be regarded as a signature of the presence of an interaction potential.\cite{valone2006ED,kraisler2013,senjean2018}  Recent work has shown that this observation can be turned to advantage in devising self-consistent exchange-correlation functionals that address the well-known derivative discontinuity and band gap problems through modified ensemble  representations.\cite{kraisler2013,senjean2018} In contrast to the expanded ladder of states that will be used to construct the atomistic potential, these electronic structure methods are designed to retain compact two-\cite{kraisler2013} and three-state\cite{senjean2018} ensemble forms, while folding in the effect of $v_{int}({\bf r})$ through renormaliztion of the contributing integer-electron densities\cite{kraisler2013} or introduction of a weight-dependent exchange-correlation functional.\cite{senjean2018}  These extensions to the isolated atom ensemble theories provide additional theoretical support for using ensemble weights to encode the effects of electronic interactions, as an alternative to determining the self-consistent densities and energies of a more restricted set of ensemble states.  

\subsubsection{The ensemble DFT atom-in-molecule decomposition}
We now summarize the ensemble DFT atom-in-molecule decomposition\cite{Atlas2020} to be used in constructing the ensemble DFT CT-EAM.

The contribution of the $i$th atom-in-molecule to the density $\rho({\bf r})$ of a larger system is expressed as dual expansions in terms of both excited state and charge state densities of isolated atoms having the same, fixed nuclear charge $Z_i$ as the atom of interest. The decomposition $\tilde{\rho}({\bf r})$ is written as:
\begin{equation}
\rho({\bf r}) \rightarrow \tilde{\rho}({\bf r}) \equiv \sum_{i=1}^{N_{at}} \rho_i^*({\bf r}),
\label{eq:DD}
\end{equation}
where the atom-in-molecule densities $\rho_i^*({\bf r})$ are given by
\begin{equation}
\rho_i^*({\bf r}) = \sum_{j= -\infty}^{Z_i - 1} \alpha_{ij} \varrho_{ij}({\bf r}),
\label{eq:AIM-expans}
\end{equation}
with weights $\alpha_{ij} \ge 0$ $\forall$ $i$,$j$, and 
the excited state ensemble densities $\varrho_{ij}({\bf r})$ for atom $i$ and charge state $j$ defined as
\begin{equation}
\varrho_{ij}({\bf r}) \equiv \sum_{k=0}^{\infty} \beta_{ijk} \rho_{ijk}({\bf r}),
\label{eq:excit-expans}
\end{equation}
with $\beta_{ijk}$ $\ge 0$ $\forall$ $i$, $j$, $k$.  The \textit{basis density} $\rho_{ijk}({\bf r})$ is thus the density of the $k$th eigenstate of the $j$th ion of atom $i$ with number of electrons $N_{ij} = Z_i + j$. $j$ labels the charge state, and $k$, the excitation state of the atom. The typical range of $j$ is between the lowest and highest oxidation states of the atom-in-molecule.  $N_i \equiv N_{i0} = Z_i$ is the number of electrons in the neutral atom.  The charge ensemble and excitation ensemble weights satisfy separate sum rules: $\sum_{j= -\infty}^{Z_i - 1} \alpha_{ij} = 1$ for each atom $i$, and $\sum_{k=0}^{\infty} \beta_{ijk} = 1$ for each atom $i$ in ionic state $j$.  

\begin{figure}[htbp]
\includegraphics[scale=.595
]{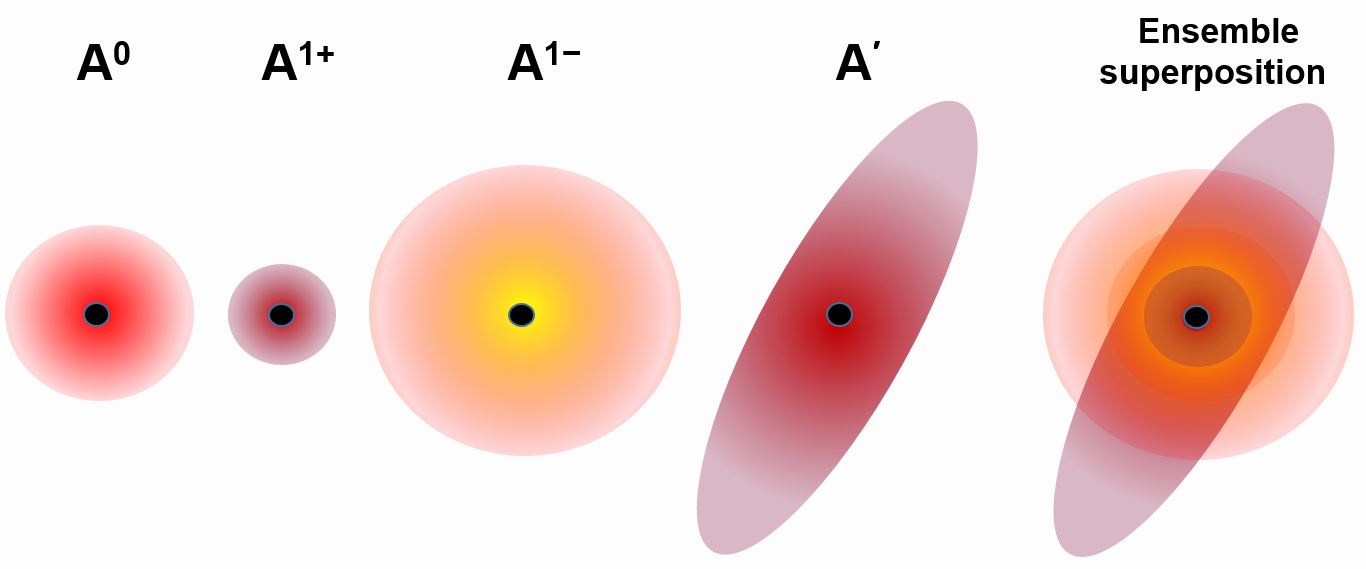}
\caption{Schematic illustration of an ensemble of electronic structure problems contributing to the characterization of an atom-in-molecule ${\rm A}^*$ with density $\rho_{\rm A}^*({\bf r})$, in a large biomolecular system. From left to right: isolated, neutral atom ground state density $\rho_{\rm A}^0({\bf r})$; isolated cation density with more compact electron density $\rho_{\rm A}^+({\bf r})$; isolated anion with diffuse electron density $\rho_{\rm A}^{-}({\bf r})$; isolated excited state density ${\rho}^{\prime}({\bf r})$ displaying charge polarization. (${\rho}^{\prime}({\bf r})$ is displayed as a non-spherical density for illustration purposes only. In practice, basis densities are spherically-averaged as noted in the text.) All four electronic structure problems derive from a common atomic potential, $v_{\rm A}({\bf r}) = -Z_{\rm A}/|{\bf r} - {\bf R}_{\rm A}|$, where ${\bf R}_{\rm A}$ is the location of nucleus A. An unnormalized equiensemble is illustrated at far right. During simulations, the ensemble weights for all states contributing to $\rho_{\rm A}^*({\bf r})$ will readjust according to chemical potential equalization (system-wide energy minimization with respect to the ensemble weights). There is no restriction on the number of states and corresponding basis densities that can be included in the ensemble for a given atom---these are chosen based on the particular integer charge and polarization states that are expected to be encountered in simulating a given system.  Excited states of ions (not illustrated) may also be included.
}
\label{fig:ESProblems}
\end{figure}

The decomposition of Eq.~(\ref{eq:DD}) can be understood intuitively from a consideration of the respective atomic and molecular potentials. Note that the external nuclear potential of the $i$th atom-in-molecule is the same for all atomic states contributing to its ensemble decomposition: $v_{i}({\bf r}) \equiv -Z_{i}/|\textbf{R}_i - \emph{\textbf{r}}|$. The decomposition uses a collection of electronic structure problems, for as many ionic and excited state excitations above the ground state as one chooses to select, to characterize the changing chemical environment of the atom-in-molecule (see Fig.~\ref{fig:ESProblems}). 
The potential for the molecule is just the sum of the individual atom-in-molecule potentials, {\it i.e.}, $v({\bf r}) \equiv \sum_{i=1}^{N_{\rm A}}  -Z_{i}/|\textbf{R}_i - \emph{\textbf{r}}|$, which is the same as the potential that would appear in the Hamiltonian for solving the exact molecular electronic structure problem.  Instead, through the dual ensemble decomposition, an approximate molecular density $\tilde{\rho}({\bf r}) \approx \rho({\bf r})$ is given by the sum of the $N_A$ ensemble atoms-in-molecule. 

The basis densities can be computed at a high level of quantum mechanical theory using standard electronic struture codes, and, as proposed elsewhere,\cite{Amokwao2020} converted to analytical radial basis functions satisfying a set of formal quantum mechanical constraints. These constraints govern the short-, medium-, and long-range behavior of atomic densities.\cite{Amokwao2020}  THis provides a further mechanism for incorporating formal chemical and physical knowledge into the structure of the force field, since it is the tails of the atom-in-molecule densities that will contribute to the embedding density in the ensemble DFT CT-EAM (see Eq.~(\ref{eq:AIM-dens}) below).

\begin{figure}[htbp]
\includegraphics[scale=.50
]{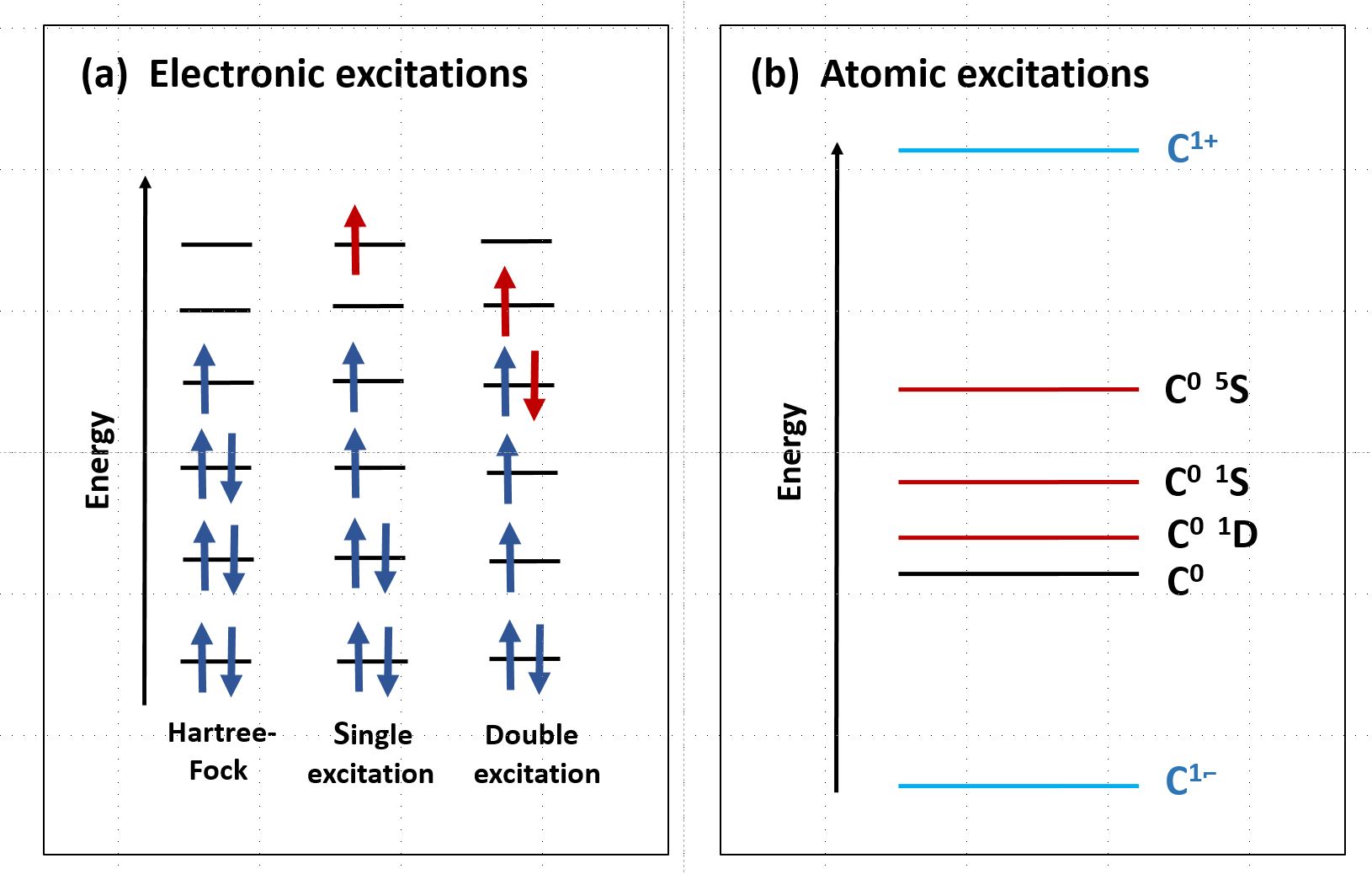}
    \caption{Particle correlations at the electronic and atomistic length scales. (a) Electrons interacting quantum mechanically. Beyond-Hartree-Fock electron correlations are computed by including determinants with increasing degrees of virtual orbital excitation within a configuration interaction (CI) expansion. Figure after Zgid {\it et al.}\cite{zgid2012} (b) Atoms interacting via an ensemble DFT classical force field.  Effects of the chemical environment on the baseline isolated atom are computed by including contributions of excited (charge-distortion) and ionic (charge-transfer) atomic states within the force field.  Illustration for carbon is schematic only (not to scale). Ionic states are indicated in blue; excited states in red. Relative energies from \citet{sasaki1974} and the NIST Atomic Spectra Database\cite{NIST2020}.}
\label{fig:lengthscales}
\end{figure}

Since the atom-in-molecule densities are expressed in terms of pre-computed, correlated basis densities, electron correlation effects are automatically propagated upward through the atom-in-molecule expansion to the atomic level. This is illustrated schematically in Fig.~\ref{fig:lengthscales}.

Note that by combining Eqs.~(\ref{eq:AIM-expans}) and (\ref{eq:excit-expans}), $\rho_i^*({\bf r})$ may also be written as
\begin{equation}
\rho_i^*({\bf r}) = \sum_{jk} w_{ijk} \rho_{ijk}({\bf r}),
\label{eq:w-wts}
\end{equation}
where $w_{ijk} \equiv \alpha_{ij} \beta_{ijk}$, and as a consequence of the $\alpha_{ij}$ and $\beta_{ijk}$ sum rules, the $w_{ijk}$ satisfy the sum rule $\sum_{jk} w_{ijk} =1$ for each $i$.

\section{Results and Discussion}
\subsection{A DFT-based charge-transfer force field}
This section presents the main result of this work, the derivation of the ensemble DFT CT-EAM from density functional theory.  The initial development and notation of Eqs.~(\ref{II.15})--(\ref{II.23}) closely follow that of Daw for the original EAM.\cite{daw1989}  We then show how to incorporate the ensemble DFT energy and atom-in-molecule representations from the previous section to derive the CT-EAM generalization.

Daw's derivation of the EAM starts from the DFT expression for the cohesive energy of a configuration $\{{\bf R}_i\}$ of $N_A$ atoms in terms of the total self-consistent electron density $\rho({\bf
r)}$: 
\begin{eqnarray}
     E_{{\rm coh}} &=& G[\rho] + \frac{1}{2}\
     {\sum_{i,j}}^{\prime}{\frac{Z_i Z_j }{|{\bf R}_i - {\bf R}_j|}} -
     \sum_i \int d{\bf r}\ {\frac{Z_i\, \rho({\bf r} ) }{|\,{\bf r} -
     {\bf R}_i |}} \nonumber \\
     &+& \frac{1}{2} \int d{\bf r}\, d{\bf r^{\prime}}\,
     {\frac{\rho({\bf r})\, \rho({\bf r^{\prime}}) }{|{\bf r} - {\bf
     r^{\prime}}|}} - \sum _i E_i.
     \label{II.15}
\end{eqnarray}
$Z_i$ and $E_i$ are the atomic number and total energy of the $i$th atom, respectively, and the prime indicates that the $i=j$ terms are omitted from the summation. $G[\rho]$ is given by:
\begin{equation}
     G[\rho] = E_{{\rm xc}} [\rho] + T_s[\rho],
     \label{eq:G}
\end{equation}
where the DFT exchange-correlation energy $E_{{\rm xc}} [\rho]$ and non-interacting kinetic energy $T_s[\rho]$ have been defined previously.\cite{fn1}  The DFT expression for the isolated-atom energies $E_i$ is:
\begin{eqnarray}
     E_i &=& E[\rho_i^a] = G[\rho_i^a] - \int d{\bf r}\ {\frac{Z_i\,
     \rho_i^a({\bf r} - {\bf R}_i) }{|\,{\bf r} - {\bf R}_i |}}
     \nonumber \\
     &+& \frac{1}{2} \int d{\bf r}\, d{\bf r^{\prime}}\,
     {\frac{\rho_i^a({\bf r} - {\bf R_i})\, \rho_i^a({\bf r^{\prime} -
     {\bf R_i}}) }{|{\bf r} - {\bf r^{\prime}}|}},
     \label{eq:isol-atom}
\end{eqnarray}
where the superscript $a$ denotes an isolated (undistorted) atomic density.

To simplify notation, it is helpful to establish the following definitions. The difference density
$\tilde{\rho}_{i}^{a}({\bf r})$ between the local electronic and
nuclear densities of the $i$th atom is defined as:
\begin{equation}
     \tilde{\rho}_{i}^{a}({\bf r})\equiv \rho _{i}^{a}({\bf r}-{\bf
     R}_{i})-Z_{i}\delta ({\bf r}-{\bf R}_{i}).
     \label{eq:diff-den}
\end{equation}
The effective classical electrostatic interaction energy  is:
\begin{equation}
U_{ij}^a \equiv \int d{\bf r}\, d{\bf r^{\prime}}\, {\frac{\tilde{\rho}_i^a(
{\bf r})\, \tilde{\rho}_j^a({\bf r^{\prime}}) }{|{\bf r} - {\bf r^{\prime}}|}
}.  \label{eq:EScl}
\end{equation}
The {\it embedding energy} ${\cal G}_{i}(\bar{\rho})$ is defined as the net exchange-correlation and non-interacting kinetic energy required to
embed the $i$th atom in an electron gas of constant density $\bar{\rho}$:
\begin{equation}
     {\cal G}_{i}(\bar{\rho})\equiv G[\rho _{i}^{a}+\bar{\rho}]-G[\rho
     _{i}^{a}]-G[\bar{\rho}] \, .
     \label{II.13}
\end{equation}
Note that electrostatic contributions are specifically excluded from the definition of ${\cal G }_{i}$, and $\bar{\rho}$ is implicitly assumed to be
neutralized by the compensating positive background density provided by a smeared-out positive background.\cite{daw1989}

It is now straightforward to show that the cohesive energy of Eq.\ (\ref{II.15}) may be rewritten in the following form:
\begin{equation}
     E_{{\rm coh}} = G[\rho({\bf r})] - \sum_i G[\rho_i^a({\bf r})] +
     \frac{1}{2} \,{\sum_{i,j}}^{\prime}U_{ij}^a.
     \label{label:Ecoh-EAM}
\end{equation}
$U_{ij}^a$ clearly corresponds to the electrostatic functions $\phi _{ij}$
in Eq.\ (\ref{eq:EAM}); approximations are required in order to identify the
analogs of the embedding functions $F_i$.

The first assumption is that the total electron density $\rho ({\bf r})$
at site $i$ can
be approximated by the sum of the undistorted atomic densities $\rho_j^a(
{\bf r} - {\bf R}_i)$:
\begin{equation}
\rho({\bf r}) \approx \sum _j \rho_j^a({\bf r} - {\bf R}_i).  \label{II.17}
\end{equation}

Note that this assumes a {\it different approximation to $\rho({\bf r})$ at each atomic site $i$}, which we indicate by $\varrho_i({\bf r})$. In constructing each local approximation, the undistorted atomic
density of atom $i$ removed, and the remainder is identified as the background density $\rho_{b,i}({\bf r})$ appropriate to that site:
\begin{equation}
     \sum _j \rho_j^a({\bf r} - {\bf R}_i) = \rho_i^a({\bf r} - {\bf
     R}_i) + \rho_{b,i}({\bf r}) \, ,
     \label{II.18}
\end{equation}
where
\begin{equation}
     \rho_{b,i}({\bf r}) \equiv \sum_{j \ne i} \rho_j^a({\bf r} - {\bf
     R}_i) \, .
     \label{II.19}
\end{equation}

The next step is to assume that the variation in $\rho_{b,i}({\bf r})$ is small in the vicinity of site $i$, relative to the dominant local
atomic density $\rho_i^a({\bf r})$, so that $\rho_{b,i}({\bf r})$ may be approximated by a constant $\bar{\rho _i}$:
\begin{equation}
     \rho_{b,i}({\bf r}) \approx \bar{\rho _i} \, .
     \label{II.19a}
\end{equation}
Eq.\ (\ref{II.19}) suggests taking
$\bar{\rho _i}$ to be the sum of electron density tails from neighboring atoms, evaluated at site $i$:
\begin{equation}
     \bar{\rho _i} \approx \rho_{b,i}({\bf r}){\big \vert}_{{\bf r} =
     {\bf R}_i} = \sum _{j \ne i} \rho_j^a({\bf R}_i - {\bf R}_j) \, .
     \label{II.19b}
\end{equation}
This is the well-known background density approximation of EAM. It is equivalent to a weighted-density approximation to $\rho_{b,i}({\bf
r})$, in the limit of delta-function localization to the nucleus\cite{daw1989}
\begin{equation}
     \bar{\rho _i} \approx \int d{\bf r} \sum_{j\ne i} \rho_j({\bf r} -
     {\bf R} _j)\ w_i({\bf r}) \, ,
     \label{II.19c}
\end{equation}
with $w_i({\bf r}) = \delta ({\bf r} - {\bf R}_i)$.

Combining Eqs.\ (\ref{II.17})--(\ref{II.19a}), the site-specific local
approximation $\varrho^{(i)} ({\bf r})$ to $\rho({\bf r})$ becomes:
\begin{equation}
     \varrho^{(i)} ({\bf r}) \approx \rho_i^a({\bf r} - {\bf R}_i) +
     \bar{\rho _i} \, .
     \label{II.19d}
\end{equation}

We now perform a series of straightforward algebraic manipulations with $G[\rho ]$, making use of Eq.\ (\ref{II.19d}), and dropping explicit spatial dependencies for clarity.  We have:
\begin{eqnarray}
     G[\rho ] &=&G[\rho ]+\sum_{i}G[\varrho^{(i)}]-\sum_{i}G[\varrho^{(i)}] \nonumber \\
     &=&\sum_{i}G[\rho _{i}^{a}+\bar{\rho _{i}}]+\left( G[\rho
     ]-\sum_{i}G[\varrho^{(i)}]\right) \nonumber \\
     &=&\sum_{i}G[\rho _{i}^{a}+\bar{\rho _{i}}]-\sum_{i}G[\bar{\rho
     _{i}}] \nonumber \\
     &\ &{\rm \hspace*{0.1in}}+\left( G[\rho ]-\sum_{i}G[\varrho^{(i)}]+\sum_{i}G[ \bar{\rho _{i}}]\right) \nonumber \\
     &=&\sum_{i}(G[\rho _{i}^{a}+\bar{\rho _{i}}]-G[\bar{\rho
     _{i}}])+E_{{\rm err} } \, ,
     \label{II.20}
\end{eqnarray}
with
\begin{equation}
     E_{{\rm err}}\equiv G[\rho ]-\sum_{i}(G[\varrho^{(i)}]-G[\bar{\rho
     _{i}}]) \, .
     \label{II.21}
\end{equation}
Subtracting $\sum_{i}G[\rho _{i}^{a}]$ from each side of Eq.~(\ref{II.20}) and using the definition of the embedding energy from Eq.~(\ref{II.13}), one finds that
\begin{equation}
     G[\rho ]-\sum_{i}G[\rho _{i}^{a}]=\sum_{i}{\cal G}_{i}[\bar{\rho
     _{i}}]+ E_{{\rm err}} \, ,
     \label{II.22}
\end{equation}
so that the cohesive energy (\ref{II.15}) becomes
\begin{equation}
     E_{{\rm coh}}=\sum_{i}{\cal G}_{i}[\bar{\rho _{i}}]+
     \frac{1}{2}\,{\sum_{i,j}}^{\prime }U_{ij}^{a}+E_{{\rm err}} \, .
     \label{II.23}
\end{equation}
This is the EAM expression derived by Daw.\cite{daw1989}  Although the focus of Daw's work was to assess the DFT-EAM connection numerically for a particular case (\textit{fcc} Ni) where electron density redistribution was expected to be small, Daw also recapitulated his derivation allowing for a charge redistribution (polarization)  $\Delta\rho_i({\bf r})$ at the location of each atom, {\it i.e.}, $\rho_i^a({\bf r}) \rightarrow \rho_i^a({\bf r}) + \Delta\rho_i({\bf r}) \equiv \varrho_i^*({\bf r})$.  He showed that an expression analogous to Eq.~(\ref{II.23}) could be derived, with $\rho_i^a({\bf r})$ replaced by $\varrho_i^*({\bf r})$ in the embedding density and electrostatic energy terms. Although the derivation does not apply to the charge-transfer case, it nevertheless presages the atom-in-molecule replacement of the static $\rho_i^a({\bf r})$ by the atom-in-molecule density $\rho_i^*({\bf r})$ introduced in the present work. 

Eq.~(\ref{II.23}) is precisely the EAM form Eq.~(\ref{eq:EAM}), with an additional error term, $E_{{\rm err}}.$  Daw further showed that under certain well-defined approximations, including the use of a local gradient approximation for the density functional $G[\rho]$, minimizing $E_{{\rm err}}$ corresponds to the determination of an optimal background embedding density. While his   derivation and analysis cannot be applied directly to situations with significant charge distortion and charge transfer, due to various localization and small-perturbation assumptions, it identified the core components that must be generalized in order to extend a DFT formulation of EAM to the strong interaction regime: (i) the density-dependent embedding functions embodying nonlocal, many-body quantum effects through their origins in the DFT functional $G[\rho]$; (ii) a pair interaction term describing electrostatic   interactions between atom-centered densities; and (iii) an embedding density expressed in terms of neighboring atom-centered densities.
 
To extend EAM to the general case we therefore propose the following mappings, starting from Eq.~(\ref{eq:EAM}) and using the basis-density ensemble expression for the atom-in-molecule density (Eq.~(\ref{eq:w-wts})) and corresponding energy ensemble weights:   
\begin{itemize}
\item[1.] Embedding function for atom $i$: $F_i \Rightarrow \sum_{M_i} \omega_{iM_i} F_{iM_i}$
\item[2.] Electrostatic energy for atom $i$: $\phi_{ij} \Rightarrow \sum_{M_iP_j} \omega_{iM_i} \omega_{jP_j} \phi_{ij,M_iP_j}$
\item[3.] Atom-centered density: $\rho_i^a({\bf r}) \Rightarrow \rho_i^*({\bf r})$
\item[4.] Embedding density: $\bar{\rho _i} \Rightarrow \bar{\rho} _i^*$
\end{itemize}
In the expressions above, since excited states of atoms with integer charge can also be included in the ensemble expansion, we have combined the excited-state and charge state indices into a single index $M_i$ looping over all $N_M$ charge and excitation ensemble states (a second combined index $P_j$ loops over the $N_P$ charge and excitation ensemble states of interacting atom $j$). With the mappings above, the final result for the general ensemble DFT CT-EAM is given by:\cite{fn8}
\begin{eqnarray}
E &=& \sum_{i} \bigg[ \sum_{M_i=1}^{N_M} \omega_{i,M_i} F_{i,M_i}
[\overline{\rho}_i^*] \nonumber \\ 
&+& \frac{1}{2}{\sum_{i,j}}^{\prime}
\sum_{M_i=1}^{N_M} \sum_{P_j=1}^{N_P} \omega_{i,M_i} \omega_{j,P_j}\Phi_{ij,M_iP_j}\bigg].
\label{eq:ensFF}
\end{eqnarray}
The background embedding electron density $\overline{\rho}_i^*$ is given by:
\begin{equation}
\overline{\rho}_i^*({\bf R}_i) \simeq \int d{\bf r}\,  \tilde{\omega}({\bf r})\, \sum_{\stackrel{j=1}{j\ne
i}}^{n_i}{\rho_j^*({\bf r}-{\bf R}_j)}.
    \label{eq:AIM-dens}
\end{equation}
$\tilde{\omega}({\bf r})$ is a weighting function that averages over the contributions of atom-in-molecule tails in a neighborhood of atom $i$.  As noted above, the standard EAM formulation assumes $\tilde{\omega}({\bf r}) = \delta({\bf r} - {\bf R}_i)$.\cite{daw1989}

Several comments are in order here.  A unique feature of the ensemble DFT CT-EAM is the requirement of internal self-consistency---between the density models appearing in the embedding and electrostatic components, and between the densities and energy functionals, which are both of ensemble form. The density consistency requirement implies that the atom-centered densities used to compute the effective background embedding density seen by the embedding functions, are identical to the atom-in-molecule densities used to compute effective electrostatic energies in the pairwise interaction term. This is because the $\rho_i^*({\bf r})$ have precise physical meaning as atom-in-molecule densities, rather than serving as mathematical proxies for a density-like quantity in a purely empirical model.  In addition, the ensemble DFT-imposed consistency requirement between the embedding energies and atom-in-molecule densities requires that their respective state-dependent ensemble weights---in the atom-in-molecule ensemble expression (\ref{eq:DD})--(\ref{eq:excit-expans}) and in the EAM expression (\ref{eq:ensFF})---correspond.         

We emphasize again that while we have presented arguments linking the various components of the EAM model---the embedding and electrostatic terms, atom-centered densities, and the embedding density---to formal DFT concepts, the relationship is not exact. In the end, Eq.~(\ref{eq:ensFF}) is an atomistic model, not a theory. The decomposition of the total density into atom-in-molecule ensembles omits interatomic correlations; the embedding density requires a weighted local averaging of a local superposition of atom-in-molecule tails; and the embedding and electrostatic interaction functions require parameterization against a set of coarse-grained physical characteristics (e.g. the energies of surfaces and defects in materials or metastable states of amino-acid conformers in proteins).  From a practical standpoint, it will be necessary to develop a stable and efficient algorithm for computing the dynamically-changing weights $\omega_{iM}$ via global chemical potential equalization.  Nevertheless, the hope is that by grounding the model in physical constructs that translate to the quantum level, it will be possible to attack a much broader range of chemical problems with greater overall fidelity than is currently feasible with more empirical approaches.

\subsection{Relation to subsystem and machine learning approaches}
The ensemble DFT formulation of Eq.~(\ref{eq:ensFF}) represents a particular approach to implementing energy-density duality within a quantum-informed force field.  In light of the Hohenberg-Kohn theorem, one way to understand this class of models is by considering how the balance between descriptions of density and energy are implemented. 

\begin{figure}
\includegraphics[scale=.19
]{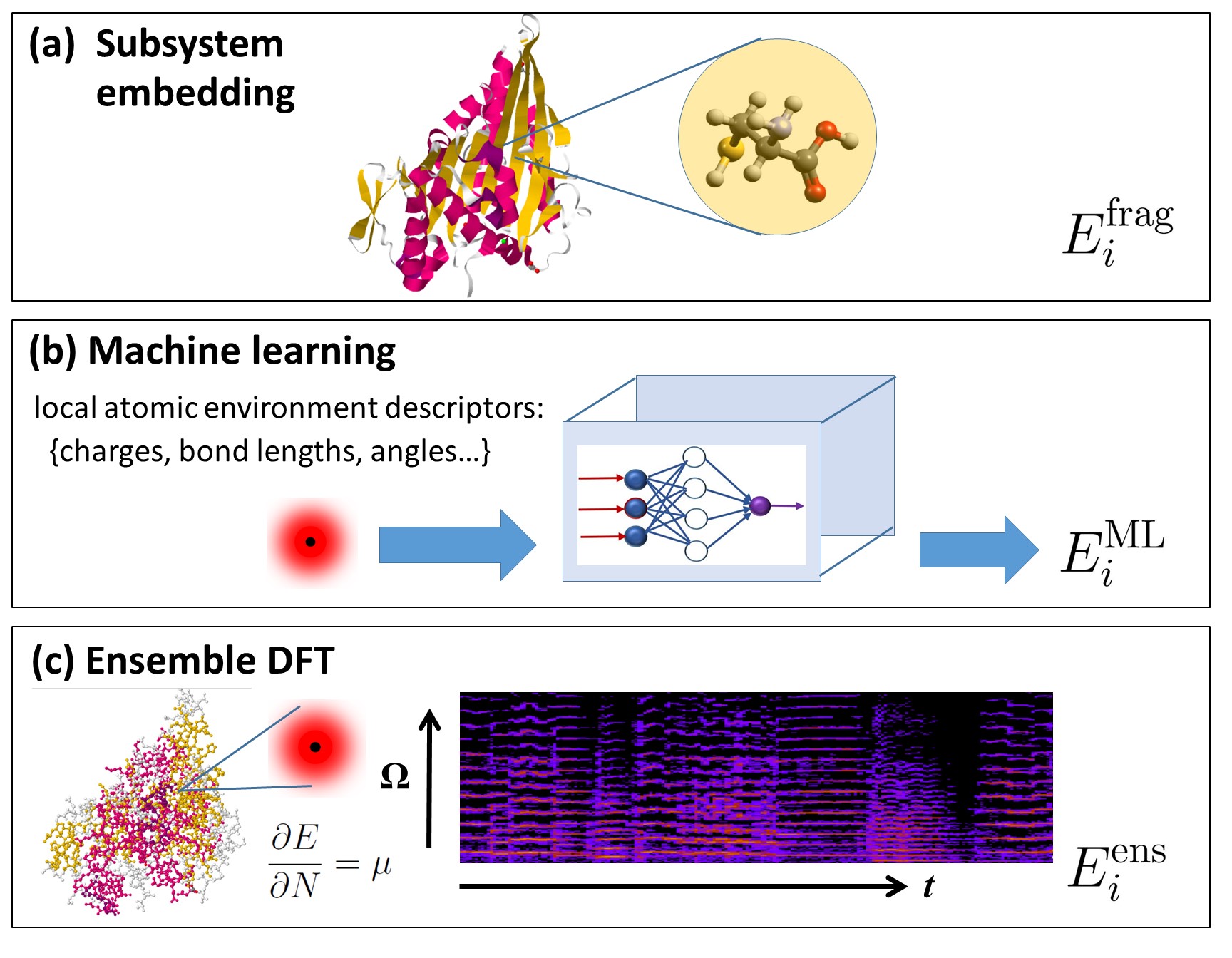}
\caption{Three approaches to energy-density duality. (a) Subsystem embedding: The subsystems of interest are embedded within the larger molecular environment. The energy $E_i^{\rm frag}$ of each subsystem---an atom or cluster of atoms---contributes to the total energy, in addition to any fragment-fragment interactions. This method can be used to study a subsystem at a higher level of theory than its surroundings, e.g. wavefunction-in-DFT. (b) Machine learning: The total energy is the sum of atomic energies computed via the application of local classifiers, typically implemented as neural networks. Environment-dependent effects and interatomic interactions are incorporated implicitly, through changes to the local descriptors input to the network as the system evolves in time. (c) Ensemble DFT CT-EAM (present work): The total energy is the sum of atomic energies computed from the ensemble DFT CT-EAM model described in the text. Relative weights of pre-computed basis densities for each atom are determined via global chemical potential equalization of an energy- and density-dependent functional. The weights determine the atomic contributions $E_i^{\rm ens}$ to the total energy.  Only the weights vary over the course of a simulation.  Here, the evolution of weights over time is represented as a notional audio spectrogram: the horizontal axis denotes time evolution of the system; the vertical axis corresponds to the discrete ensemble energy states of each atom in the system (neutral, ionic, excited, labeled by frequency $\Omega$), and color corresponds to the amplitude (ensemble weight) of a given energy state's contribution to the total. Image credits: (a) and (c), JSMol visualizations of kinesin-1 motor domain, PDB ID 5LT0\cite{cao2017}; (c), illustrative spectrogram generated using https://academo.org/demos/spectrum-analyzer/.}
\label{fig:energy-density}
\end{figure}

For comparison, the two methods most closely related to the present approach---subsystem embedding and machine learning---are illustrated in Fig.~\ref{fig:energy-density}, together with ensemble DFT CT-EAM.  

In all three approaches, there is a notion of quantum mechanical embedding of a molecular subsystem, be it a cluster or a single atom, within a broader chemical environment. Importantly, this chemical environment is also modeled quantum mechanically, and regardless of the details of the embedding process, there is a mechanism for quantum mechanical registration between the subsystem(s) and their environment. In the present approach, this is accomplished by imposing \textit{theoretical} self-consistency between the ensemble density and ensemble energy representations, as in the earlier isolated-atom ensemble DFT formulations, and \textit{computational} self-consistency, equivalent to global energy minimization through chemical potential equalization,\cite{sanderson1951,parr1978,rappe1991,rick1994} for determining the dynamically-changing ensemble weights.  In all three models, the total energy is expressed as the sum of component energies (for subsystem DFT there are additional contributions from coupling to the surrounding environment; for ensemble DFT there is a an additional summation over pairwise electrostatic interactions).  

It is instructive to further compare the two atom-centric methods---machine learning and ensemble DFT CT-EAM---to appreciate the conceptual differences in their formulations, and identify potential opportunties for cross-fertilization between their respective data-driven and theory-driven approaches. 

{\it The (meta-)density as descriptor.} In keeping with energy-density duality, the present work utilizes a physical atom-in-molecule density as input to the energy functional of each atom. In the first generation of neural-network potentials, the emphasis was on demonstrating the feasibility of data-driven, statistical machine learning approaches for reproducing quantum-computed energies, or even improving upon them.\cite{ramakrishnan2015} Input descriptors were therefore constructed based on local geometry and modeling intuition, and descriptor and (neural network) classifier designs emphasized the use of mathematical basis functions to facilitate adherence to symmetry constraints.\cite{behler2007,behler2011} 

As a result, many of these atomic descriptor models can now be seen as specific instances of a common, abstract structural representation.\cite{bartok2013,willatt2019}  This has led to the development of the bispectrum family of descriptors\cite{bartok2013} and related potentials.\cite{bartok2010,thompson2015,wood2019} In parallel with these advances, and motivated by the goal of generating potential energy surfaces with quantum-level accuracy, there is growing interest in developing more realistic models of the electron-density-as-descriptor.\cite{ghasemi2015,grisafi2018, sinitskiy2018,dick2019} These include DFT \cite{bogojeski2020} and machine-learned ``atom-in-molecule''\cite{fn5} models.\cite{zubatyuk2019,huang2020} Such data-driven approaches are complementary to the physical modeling of the atom-in-molecule described here. A potentially fruitful avenue for future research will be to explore combinations of these very different perspectives.

{\it Energy-density duality and model generalizability.}  Any theoretical model that is not exact introduces implicit bias into its predictions.  In DFT, the bias enters in the form of approximations to the exchange-correlation functional: some functionals are oriented toward accurate descriptions of molecular structure; others, the study of chemical reactions; and still others, the study of extended systems (including  functionals tailored for various sub-categories: metals, insulators, strongly-correlated materials, etc.)  These biases are mirrored in the design of force field models, where they manifest themselves in the form of energy-density duality.\cite{fn4}  However cleverly a model is designed, there are inevitable tradeoffs in how particular physico-chemical effects are built into model input (density, density descriptor vector) or energy predictor (model functional, machine learning classifier). Indeed, the act of model parameterization not only sets the values of hyperparameters, it also serves to compensate for missing correlations in the structure of the model itself.\cite{valone2006Kos} That is why physically-motivated, compact representations such as the one described here have a potential advantage over purely mathematical models containing thousands of parameters.  In complex systems where competing chemical environments often co-exist (for instance, a solvated protein interacting with a surface), achieving a good balance between energy and density models, and the number of model parameters, can have a significant impact on generalizability.

In the present approach, the propagation of information from the quantum to atomistic length scale illustrated in Fig.~\ref{fig:lengthscales} is not perfect: there is missing interatomic correlation due to the approximate  nature of the ensemble atom-in-molecule decomposition and corresponding ensemble representation of the energy.  To preserve the simplicity of the model and avoid the need to parameterize angular-dependent basis densities, the atom-in-molecule basis densities are sphericalized:\cite{Amokwao2020} theoretical rationales for this strategy come from DFT\cite{theophilou2018} and the theory of function approximation. Nevertheless, it will be important to explore systematically the extent to which the ensemble representation, in combination with introduction of the charge- and polarization-dependent embedding energy functionals, is able to compensate for this missing information within the force field.

{\it Energy-density duality and length scale coupling.} The $\Delta$-DFT approach,\cite{bogojeski2020} motivated by the work of \citet{ramakrishnan2015}, uses a two-level machine learning approach to first learn a coupled-cluster-accuracy electron density from an input DFT density, followed by a second machine learning model which uses learned DFT densities to predict the total energy. This approach is of particular interest in the context of the present work, since it ties the construction of a machine-learned energy functional designed for atomistic simulation, to a machine-learned model for the true electron density functioning as a chemical descriptor, and links the two through the medium of DFT.  A second notable feature, as discussed further below, is the explicit recognition of the importance of incorporating a range of excitation energies above the ground state in designing and training the density representation against a conformer training set.   

{\it Reference states and training data.}  We have previously noted an analogy between the zero-temperature electronic excitations characterizing spatial electron correlations, and atomic charge state excitations characterizing dynamic  correlations among atoms interacting in a finite-temperature MD simulation.  Correlations-as-electronic excitations are mirrored upward in length scale, to become correlations-as-atomic-charge-and-excited-states. The construction of the DFT ensemble force field makes explicit use of this analogy. At the next length scale up, the existence of stable and metastable structures that are free-energy-accessible at finite temperature can inform the selection of reference states and configurations for parameterization.  

Our goal in this work has been to present a principled, DFT-based force field that can help explore not only the known states of the system, but also discover and explore the unexpected ones.  An important application will be the construction of kinetic models based on quantum-informed, reactive-MD explorations of biomolecular potential energy surfaces.\cite{klus2018} In order to discover new states contributing to mechanistic dynamics, it is essential to be able to sample all relevant (and potentially unanticipated) atomic excitations arising from chemical interactions and reactive dynamics. In the machine learning case, this translates to the question of generalizability, and whether the force field can accurately extrapolate predictions from a limited dataset of randomly-sampled chemical bonding environments. For the ensemble DFT CT-EAM force field, the ensemble representation will guide the choice of parameterization strategy to emphasize the selection of a compact set of physically- and chemically-motivated structures.

\section{Conclusion}
In the language of modern machine learning, DFT is Natures's classifier for electronic structure.  By formulating the CT-EAM in terms of ensemble DFT, we have designed a principled force field that describes both polarization and charge-transfer excitations relative to the ground states of the interacting atoms.  The respective excited state and ionic charge densities further function as Nature's descriptors for characterizing local changes in electronic structure due to each atom's dynamically-changing environment. In this way, the ensemble-based atom-in-molecule formulation has a direct translation, through density functional theory, from the electronic to the atomic level. The coupling between length scales is mediated by the electron density---a quantum mechanical scalar field that is also an experimental observable.  The new force field constitutes an atom-in-molecule ensemble charge-transfer generalization of the original embedded atom method (EAM) force field,\cite{daw1983,daw1984} originally developed for atomistic simulations of elemental materials assuming fixed atomic density distributions.\cite{daw1989}  The foundation in EAM makes it possible to take advantage of a rich body of experience from condensed phase modeling: for example, information-theoretic\cite{baskes1987,muralidharan2007ED} and physics-based\cite{valone2014,valone2016} functional forms for the embedding energy; strategies for the judicious selection of structural reference systems\cite{valone2006Kos,valone2014,valone2016} to be used in parametrization, such as metastable (excited-state) amino acid conformers,\cite{Amokwao2012,monti2013} structures derived from symmetric dilation of vibrational modes,\cite{muralidharan2007ED,muralidharan2007Si} ionic fragments of mechanistic importance for a given system (e.g. ${\rm OH}^-$, ${\rm H}_3{\rm O}^+$, ${\rm H}_5{\rm O}_2^+$ and ${\rm H}_9{\rm O}_4^+$ for water);\cite{muralidharan2007ED,schmitt1999,day2002} and chemical insights derived from the wavefunction-based fragment Hamiltonian (FH) approach for characterizing atomic charge-transfer excitations within the embedding function.\cite{valone2011IM,valone2011WF,valone2014,valone2016}  The atom-in-molecule formulation in terms of quantum-mechanical, physics-constrained atomic and ionic basis densities\cite{Amokwao2020} enables a straightforward interpretation of the dynamically-evolving chemistry effected through chemical potential equalization of ensemble weights.  This is in contrast to machine-learned quantum chemical force fields, where it can be difficult to tease apart descriptor and descriptor-energy redundancies, and the use of automated descriptor discovery or empirically-chosen functions present ongoing challenges to interpretability.\cite{du2019} 

The rigorous foundation of the force field in DFT, and its formulation in terms of atoms rather than larger fragments, affords a number of key advantages.  While this work was  motivated by the need for scalability to large, heterogeneous systems with many different elements, and the study of reactive biomolecular chemistry, it is equally applicable to materials problems such as charge transfer in the vicinity of interfaces and for complex alloys.\cite{streitzPRB1994,haftel2001,zhou2004}  Since the characteristics of the local chemical environment are  encoded through an atom's canonical set of basis densities and embedding energy functions, there is no need for multiple ``oxygens'' or ``carbons'' in order to parameterize and apply a force field for a particular atom under differing chemical conditions: the set of basis states is unique and immutable for a given element.  
The foundation in DFT also provides a mechanism for a principled coupling between the quantum mechanical, atomistic, and kinetic modeling regimes, through the determination of Markov states extracted from chemically-informed MD trajectories.\cite{noe2017,klus2018} Carrying out such a multiscale programme may enable simulations to finally reach biophysically-relevant timescales without the loss of
quantum- and atomic-level details critical to describing charge-transfer-mediated (catalytic) dynamics.  We are currently proceeding to applications of the ensemble DFT CT-EAM force field to the energetics of the amino acids, as fundamental building blocks for the study of proteins and their interactions, and as the foundation for constructing chemically-informed kinetic models of protein dynamics. 

\begin{acknowledgements}
It is a pleasure to acknowledge Steve Valone for 
numerous stimulating and insightful discussions of this problem. The author is grateful to the members of her research group, past and present, for their valuable contributions. 
This work was supported in part by the National Science Foundation and by the DoD/DTRA CB Basic Research Program (Grant No. HDTRA1-09-1-0018).
\end{acknowledgements}

\bibliography{ensemble.bib}
\end{document}